\documentclass[
    ,final            
  ]
  {aipproc}

\layoutstyle{8x11single}

\def\lsim{\raise0.3ex\hbox{$<$\kern-0.75em\raise-1.1ex\hbox{$\sim$}}}
\def\gsim{\raise0.3ex\hbox{$>$\kern-0.75em\raise-1.1ex\hbox{$\sim$}}}

\def\mean#1{\left<#1\right>}
\def\Journal#1#2#3#4{ {\emph{#1}} {\textbf{#2}}, #3 (#4)}

\def\NPA{{Nucl. Phys.}~{\rm A}}

\def\PLB{{Phys. Lett.}~{\rm B}}

\def\PRL{Phys. Rev. Lett.\ }
\def\PRD{{Phys. Rev.}~{\rm D}}
\def\PRC{{Phys. Rev.}~{\rm C}}

\setcitestyle{numbers}
\begin{document}

\title{Fragmentation Functions in-Medium, Two Particle Correlations and Jets in PHENIX at RHIC}

\keywords      {Fragmentation Functions, Correlations, Jets, {Supported by the U.S. Department of Energy, Contract No. DE-AC02-98CH1-886.} 
}
\author{M.~J.~Tannenbaum for the PHENIX Collaboration}{
  address={Physics Department, 510c, Brookhaven National Laboratory, Upton, NY 11973-5000, USA}}



\begin{abstract}
Latest results from the PHENIX experiment at RHIC on these topics will be presented. 
Results will be shown for Au+Au compared to p-p collisions as well as compared to results from fully reconstructed jets at LHC. \end{abstract}

\maketitle

\begin{figure}[!bh]
\includegraphics[height=0.25\textheight]{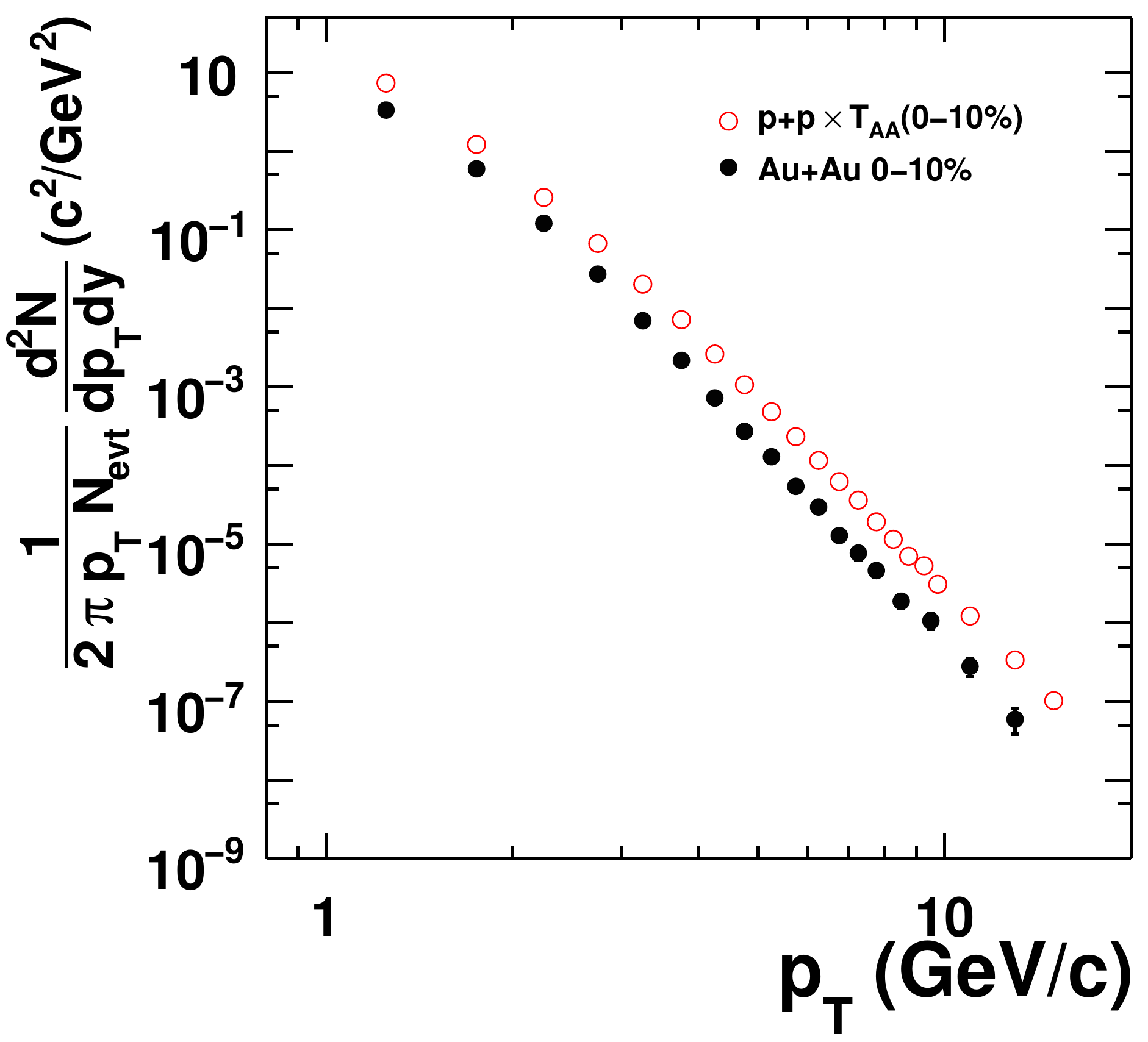}
\hspace*{0.04\textwidth} \includegraphics[height=0.25\textheight]{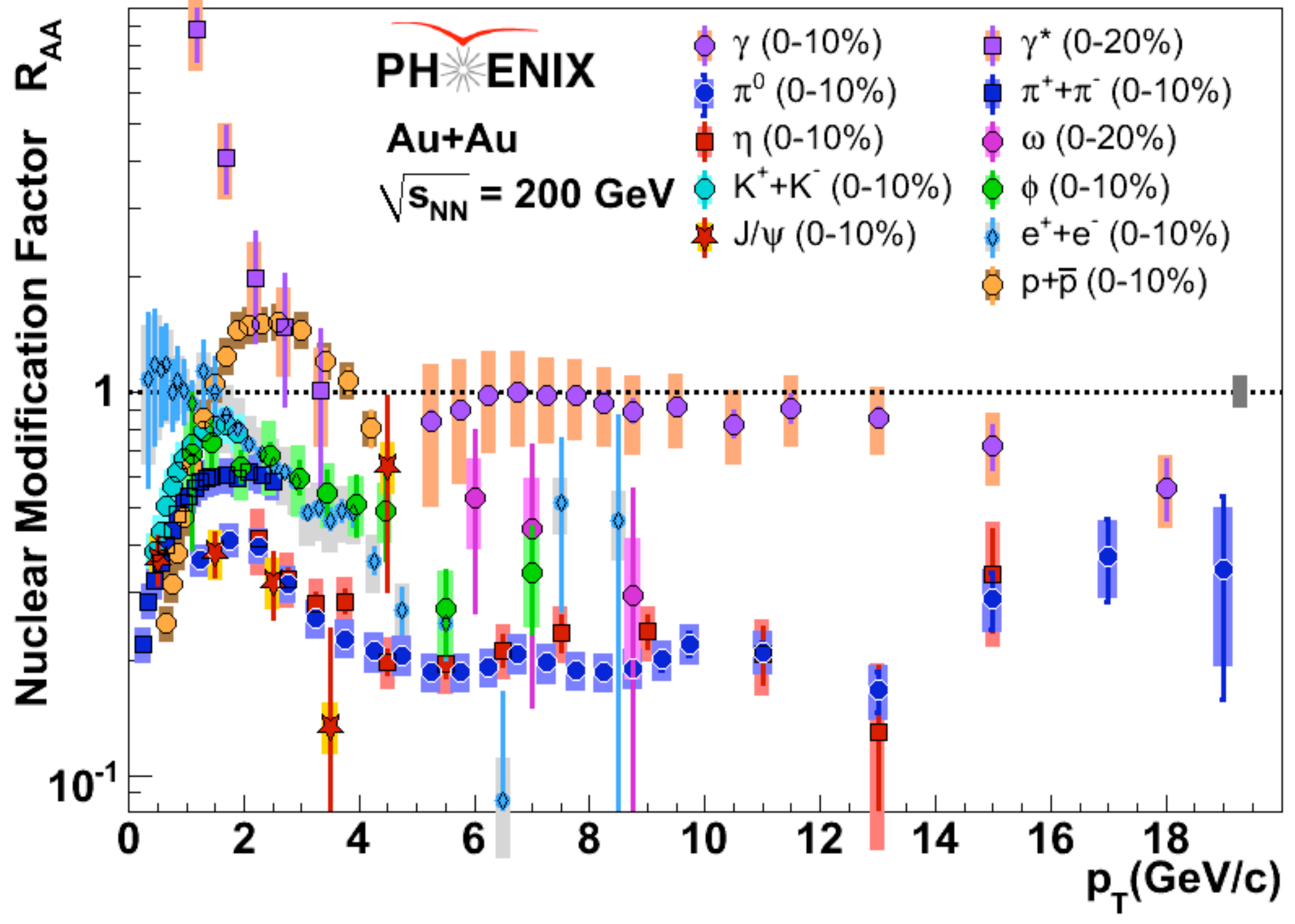}
\caption{a) (left) Invariant yield of $\pi^0$ at $\sqrt{s_{NN}}=200$ GeV as a function of transverse momentum $p_T$ in p-p collisions multiplied by $\mean{T_{AA}}$ for Au+Au central (0--10\%) collisions compared to the Au+Au measurement~\cite{ppg054}. b) (right) $R_{AA}(p_T)$ for all identified particles so far measured by PHENIX in Au+Au central collisions at $\sqrt{s_{NN}}=200$ GeV.}
\label{fig:Tshirt}
\end{figure}

   The discovery, at RHIC, that $\pi^0$ produced at large transverse momentum, $p_T$, from hard-scattering of the constituent partons of the nucleons in nuclei are suppressed in central Au+Au collisions by roughly a factor of 5 compared to point-like scaling from p-p collisions is arguably {\em the}  major discovery in Relativistic Heavy Ion Physics. For $\pi^0$ (Fig.~\ref{fig:Tshirt}a)~\cite{ppg054} the hard-scattering in p-p collisions is indicated by the power law behavior $p_T^{-n}$ with $n=8.1\pm 0.05$ for the invariant cross section, $E d^3\sigma/dp^3$, for $p_T\geq 3$ GeV/c. The Au+Au data can be characterized either as shifted down in energy relative to the point-like scaled p-p data, or down in magnitude, i.e. suppressed. In Fig.~\ref{fig:Tshirt}b, the suppression of the many identified particles measured by PHENIX at RHIC is presented as the Nuclear Modification Factor, 
$R_{AA}(p_T)$, the ratio of the yield of  per central Au+Au collision (upper 10\%-ile of observed multiplicity)  to the point-like-scaled p-p cross section:
   \begin{equation}
  R_{AA}(p_T)=[{{d^2N^{\pi}_{AA}/dp_T dy N_{AA}}]/ [{\langle T_{AA}\rangle d^2\sigma^{\pi}_{pp}/dp_T dy}}] \quad , 
  \label{eq:RAA}
  \end{equation}
where $\mean{T_{AA}}$ is the overlap integral of the nuclear thickness functions. 
The striking differences of $R_{AA}(p_T)$ in central Au+Au collisions for the many particles measured by PHENIX  (Fig.~\ref{fig:Tshirt}b) illustrates the importance of particle identification for understanding the physics of the medium produced at RHIC. Most notable are the equal suppression by a constant factor of 5 of $\pi^0$ and $\eta$ for $4\leq p_T \leq 15$ GeV/c, the equality of suppression of direct-single $e^{\pm}$ (from heavy quark ($c$, $b$) decay) and $\pi^0$ at $p_T\gsim 5$ GeV/c, the non-suppression of direct-$\gamma$ from the QCD Compton effect~\cite{QCDcompton}, $g+q\rightarrow \gamma+q$, for $p_T\geq 4$ GeV/c and the exponential rise of $R_{AA}$ of direct-$\gamma$ for $p_T<2$ GeV/c~\cite{ppg086}, which is totally and dramatically different from all other particles and attributed to thermal photon production by many authors (e.g. see citations in reference~\cite{ppg086}). The fact that all hadrons are suppressed for $p_T\gsim 5$ GeV/c, but direct-$\gamma$ are not suppressed, indicates that suppression is a medium effect on outgoing color-charged partons.

The key to measuring the fragmentation function of the jet of particles from an outgoing hard-scattered parton  is to know the energy of the original parton which fragments, as pioneered at LEP~\cite{Tingpizff}. Thus, a measurement of the direct-$\gamma-h$ correlation from $g+q\rightarrow \gamma+q$, where the $h$ represents charged hadrons opposite in azimuth to the direct-$\gamma$,  is (apart from the low rate) excellent for this purpose since both the energy and identity of the jet (8/1 $u$-quark, maybe 8/2 if the $\bar{q}+q\rightarrow \gamma+g$ channel is included) are known to high precision. 
		Two particle correlations are analyzed in terms of the two variables~\cite{ppg029}: $p_{\rm out}=p_T \sin(\Delta\phi)$, the out-of-plane transverse momentum of a track;  
 and $x_E$, where:\\ 
\begin{minipage}[c]{0.5\textwidth}
\vspace*{-0.30in}
\begin{equation}	
x_E=\frac{-\vec{p}_T\cdot \vec{p}_{Tt}}{|p_{Tt}|^2}=\frac{-p_T \cos(\Delta\phi)}{p_{Tt}}\simeq \frac {z}{z_{\rm trig}}  
\label{eq:mjt-xE}
\end{equation}
\vspace*{0.06in}
\end{minipage}
\hspace*{0.01\textwidth}
\begin{minipage}[b]{0.50\textwidth}
\vspace*{0.06in}
\includegraphics[scale=0.6]{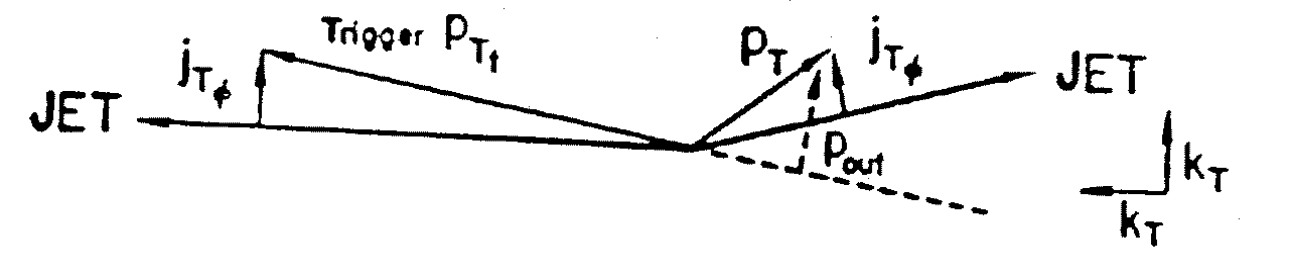}
\vspace*{-0.12in}
\label{fig:mjt-poutxe} \nonumber
\end{minipage}

\noindent $z_{\rm trig}\simeq p_{Tt}/p_{T{\rm jet}}$ is the fragmentation variable of the trigger jet, and $z$ is the fragmentation variable of the away jet. Note that $x_E$ would equal the fragmenation fraction $z$ of the away jet, for $z_{\rm trig}\rightarrow 1$, if the trigger and away jets balanced transverse momentum. The beauty of direct-$\gamma$ for this purpose is that $z_{\rm trig}\equiv 1$. 

\begin{figure}[h]
\includegraphics[height=0.21\textheight]{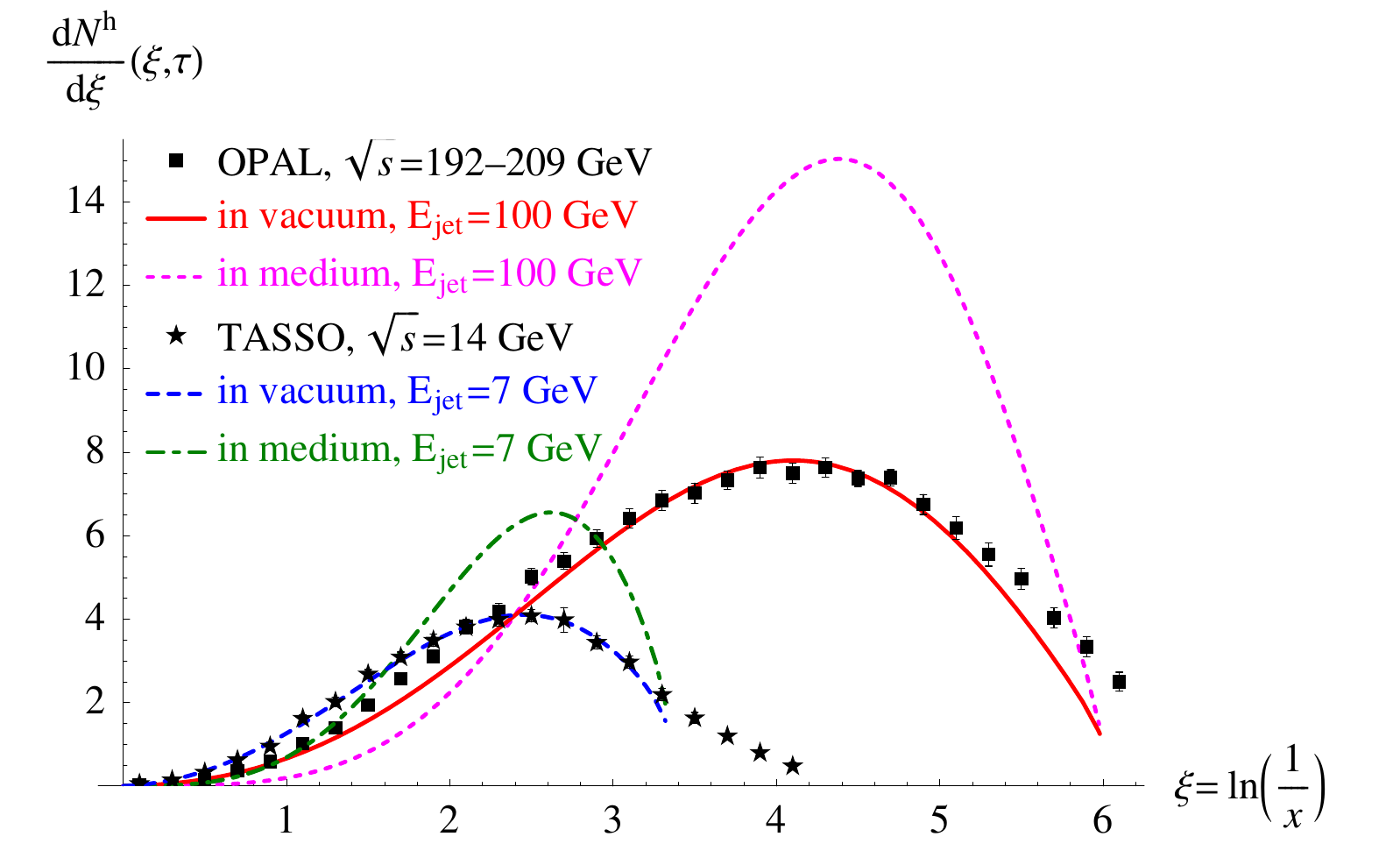}
\includegraphics[height=0.20\textheight]{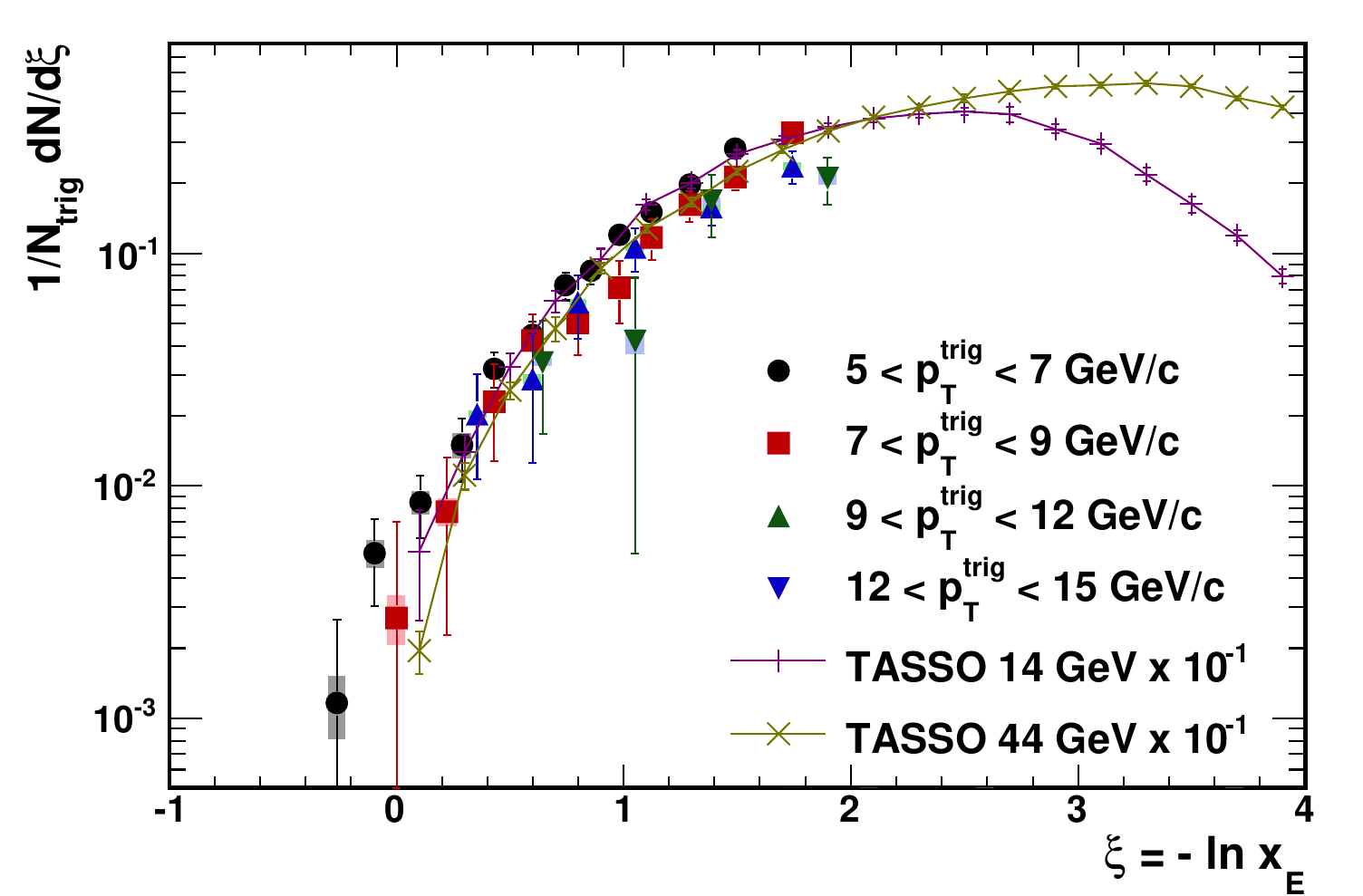}
\caption{a) Predicted $\xi$ distributions in vacuum and in medium for two jet energies~\cite{BW06}, together with measurements from $e^+ e^-$ collisions.   b) $\xi=-\ln\, x_E$ distributions for PHENIX isolated direct-$\gamma$ data~\cite{ppg095} for all $p_{T_t}$ ranges combined, compared to $e^+ e^-$ collisions at $\sqrt{s}=14$ and 44 GeV.}
\label{fig:BorgWied}
\end{figure}

\begin{figure}[!bh]
\includegraphics[height=0.25\textheight]{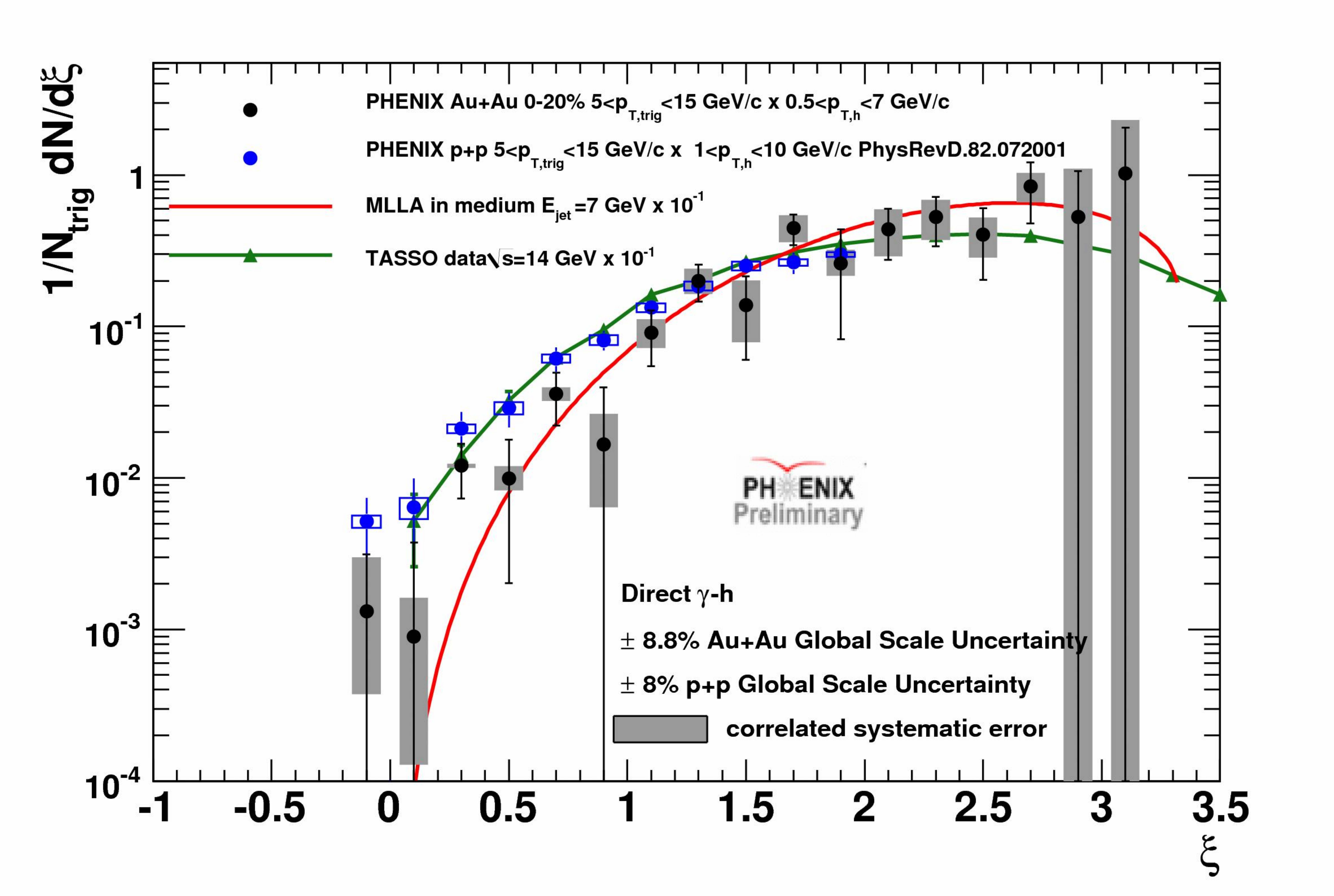}
\includegraphics[height=0.24\textheight]{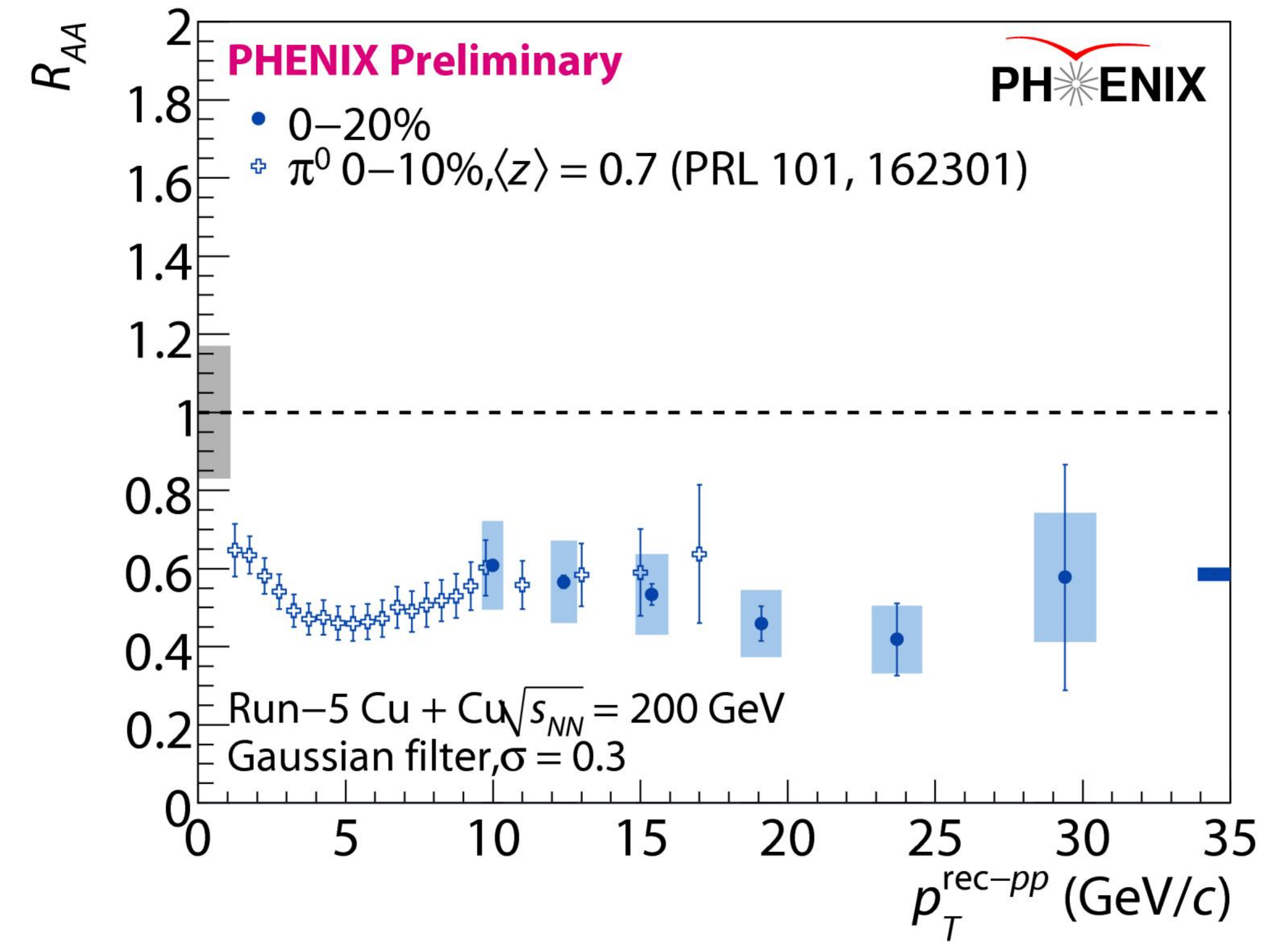}
\caption{PHENIX Preliminary results: a) (left) $\xi=-\ln\, x_E$ distribution in Au+Au (0--20\%)~\cite{ConnorsHP2010}; b) (right) $R_{AA}(p_T)$ for jets compared to $\pi^0$ in Cu+Cu (0--20\%) central collisions~\cite{LaiHP2010}. See references~\cite{ConnorsHP2010,LaiHP2010} for details.}
\label{fig:AA-ffJet}
\end{figure}

Following the approach of
Borghini and Wiedemann (Fig.~\ref{fig:BorgWied}a)~\cite{BW06} who predicted the medium modification of fragmentation functions in the hump-backed or $\xi=\ln(1/z)$ representation, PHENIX measured $x_E$ distributions and converted to the $\xi=-\ln\, x_E$ representation as shown in Fig.~\ref{fig:BorgWied}b~\cite{ppg095} which are in quite excellent agreement with the dominant $u$-quark fragmentation functions measured in $e^+ e^-$ collisions at $\sqrt{s}/2=7$ and 22 GeV, which cover a comparable range in jet energy. 
PHENIX preliminary measurements~\cite{ConnorsHP2010} of the $\xi=-\ln\, x_E$ distribution in central (0--20\%) Au+Au collisions, which suggest a modification consistent with Ref.~\cite{BW06} are shown in Fig.~\ref{fig:AA-ffJet}a. Final results with improved statistics should be available shortly. Fragmentation functions from full jet reconstruction in A+A collisions are not yet available although analysis has reached the stage of preliminary 
results for $R_{AA}(p_T)$ from jets in Cu+Cu collisions (Fig.~\ref{fig:AA-ffJet}b)~\cite{LaiHP2010}. These results agree well with the $\pi^0$ measurements but, importantly, go to much larger $p_T$, which is the main advantage, so far. 

\begin{figure}[!bh]
\includegraphics[height=0.25\textheight]{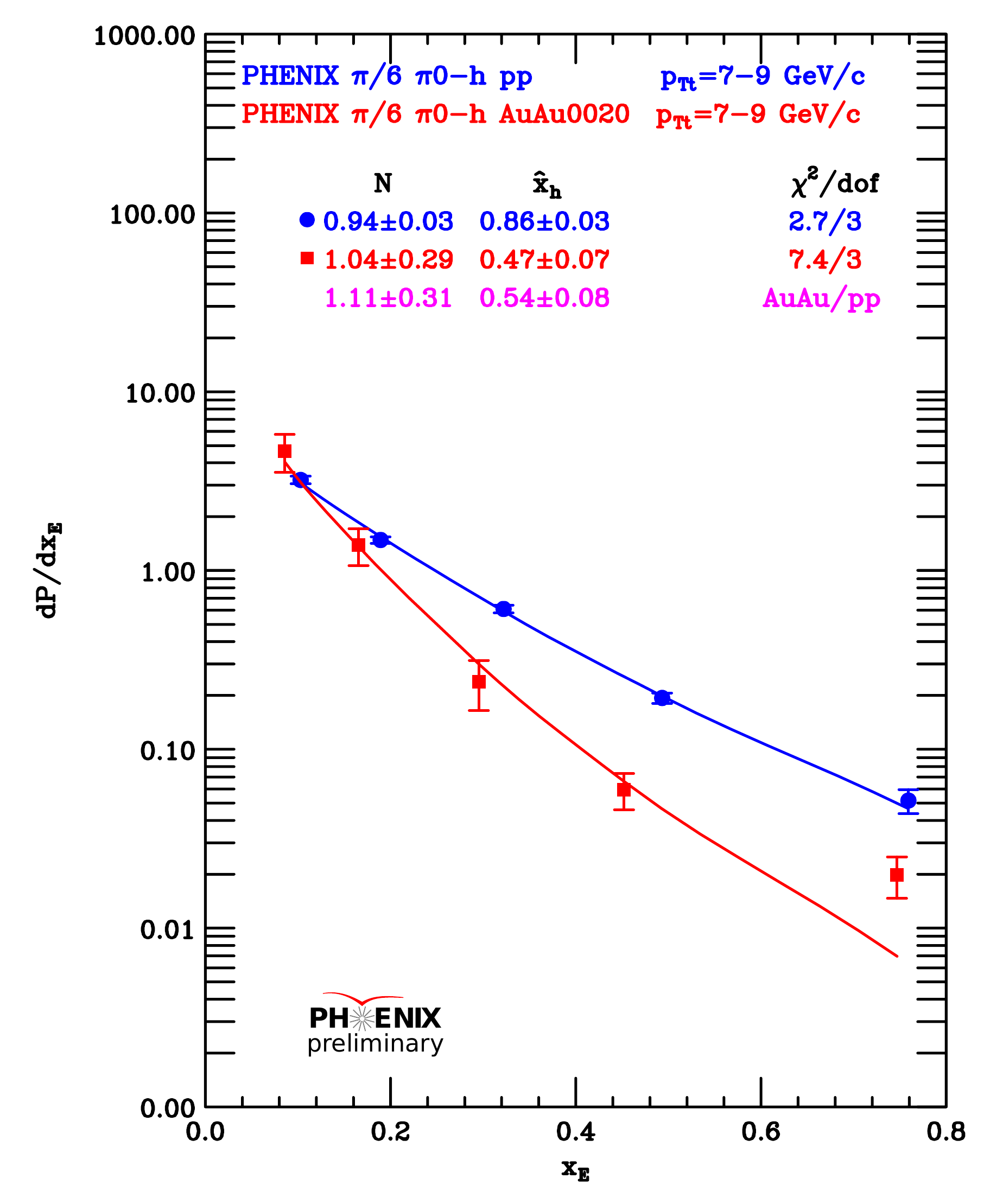}
\includegraphics[height=0.25\textheight]{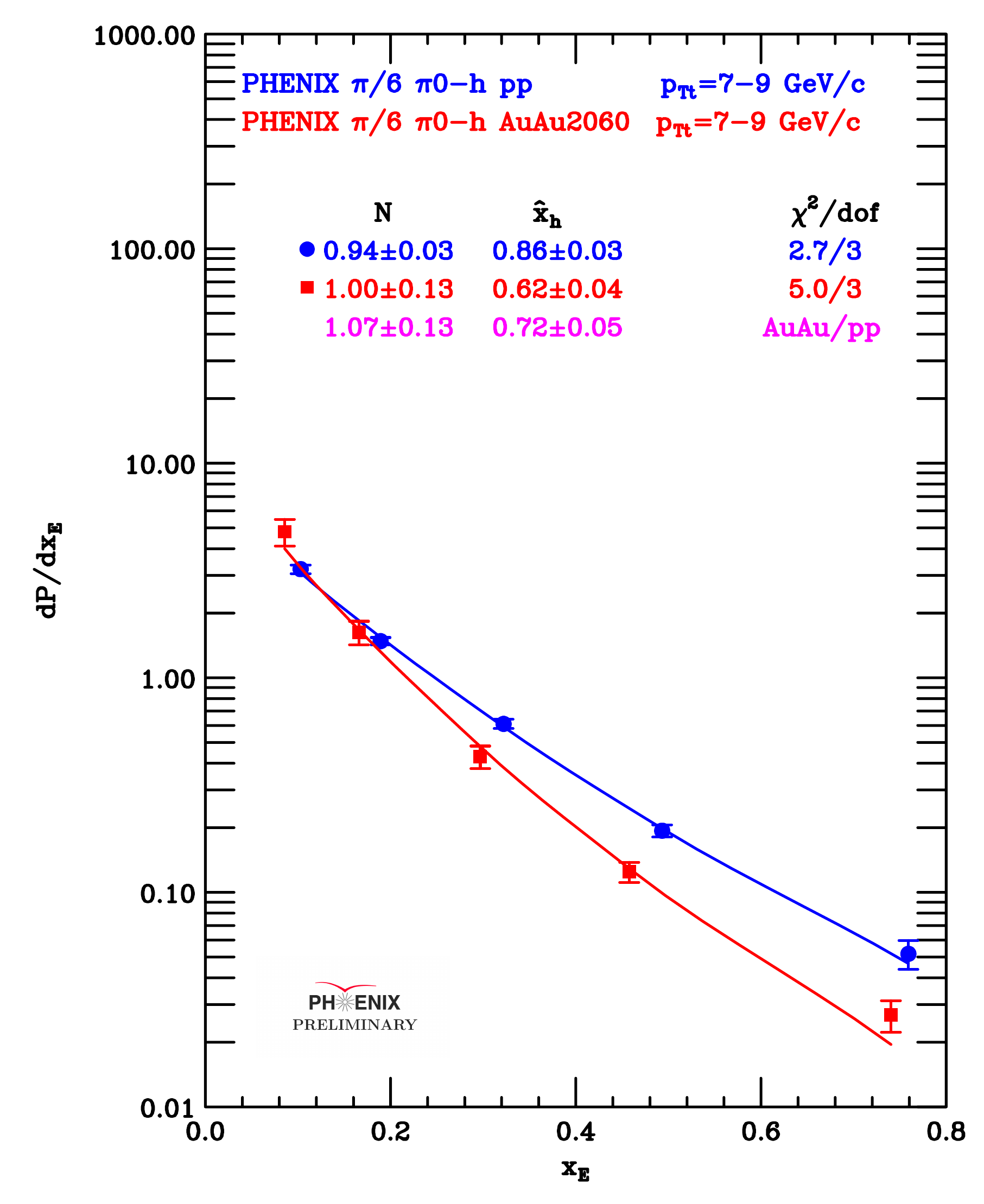}
\includegraphics[height=0.25\textheight]{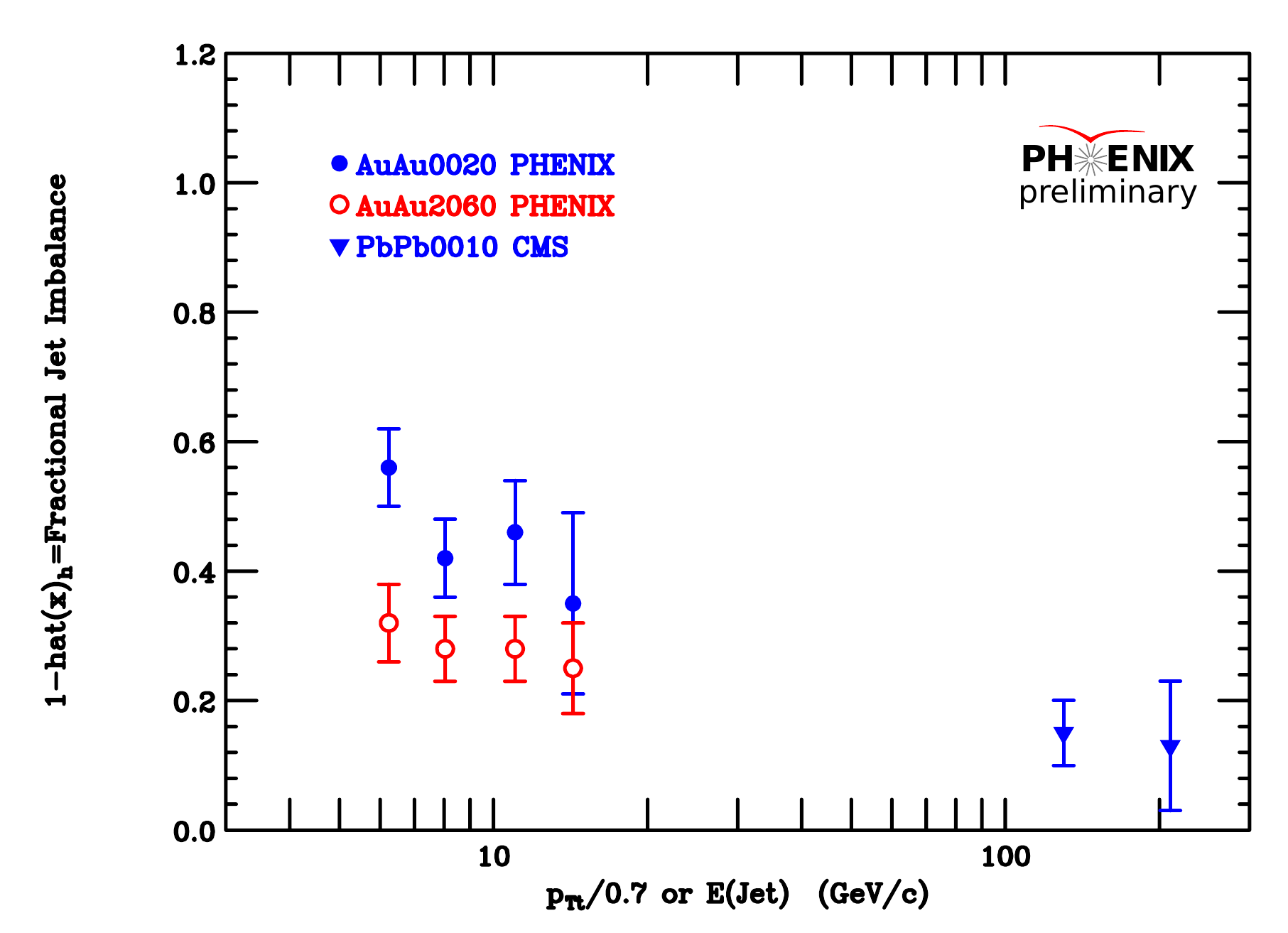}
\caption{(left) $x_E$ distributions~\cite{MJT-Utrecht} from p-p (blue circles) and AuAu  (red squares) for $p_{T_t}=7-9$ GeV/c, together with fits to Eq.~\ref{eq:condxeN2} p-p (solid blue line), AuAu (solid red line) with parameters indicated: a) 00-20\% centrality, b) 20--60\% centrality. The ratios of the fitted parameters for AuAu/pp are also given. c) (right) Fractional jet imbalance~\cite{MJT-Utrecht}, $1-\hat{x}_h^{AA}/\hat{x}_h^{pp}$, for RHIC and CMS data. }
\label{fig:AuAupp79}
\end{figure}

One of the important lessons learned about fragmentation functions at RHIC~\cite{ppg029} is that the away-side $x_E$ distribution of particles opposite to a trigger particle (e.g. a $\pi^0$), which is itself the fragment of a jet, does not measure the fragmentation function, but, instead, measures the ratio of $\hat{p}_{T_a}$ of the away-parton to $\hat{p}_{T_t}$ of the trigger-parton and depends only on the same power $n$ as the invariant single particle spectrum:  
		     \begin{equation}
\left.{dP \over dx_E}\right|_{p_{T_t}}\approx {N\,(n-1)}{1\over\hat{x}_h} {1\over
{(1+ {x_E \over{\hat{x}_h}})^{n}}} \, \qquad . 
\label{eq:condxeN2}
\end{equation}
This equation gives a simple relationship between the ratio, $x_E\approx p_{T_a}/p_{T_t}\equiv z_T$, of the transverse momenta of the away-side particle to the trigger particle, and the ratio of the transverse momenta of the away-jet to the trigger-jet, $\hat{x}_{h}=\hat{p}_{T_a}/\hat{p}_{T_t}$. PHENIX measurements~\cite{ppg106} of the $x_E$ distributions of $\pi^0$-h correlations in p-p and Au+Au collisions at $\sqrt{s_{NN}}=200$ GeV were fit to Eq.~\ref{eq:condxeN2} (Fig.~\ref{fig:AuAupp79}a,b)~\cite{MJT-Utrecht}.
The results for the fitted parameters are shown on the figures. In general the values of $\hat{x}^{pp}_h$ do not equal 1 but range between $0.8<\hat{x}^{pp}_h<1.0$ due to $k_T$ smearing and the range of $x_E$ covered. In order to take account of the imbalance ($\hat{x}^{pp}_h <1$) observed in the p-p data, the ratio $\hat{x}_h^{AA}/\hat{x}_h^{pp}$ is taken as the measure of the energy of the away jet relative to the trigger jet in A+A compared to p-p collisions. The fractional jet imbalance was also measured directly with reconstructed di-jets by the CMS collaboration at the LHC in Pb+Pb central collisions at $\sqrt{s_{\rm NN}}=2.76$ TeV~\cite{CMSdijet}; and with the large effect in p-p collisions corrected in the same way~\cite{MJT-Utrecht}, the results compared to PHENIX are shown in Fig.~\ref{fig:AuAupp79}c. The large difference in fractional jet imbalance between RHIC and LHC c.m. energies could be due to the difference in jet $\hat{p}_{T_t}$ between RHIC ($\sim 20$ GeV/c) and LHC ($\sim 200$ GeV/c), the difference in $n$ for the different $\sqrt{s}$, or to a difference in the properties of the medium. Future measurements will need to sort out these issues by extending both the RHIC and LHC measurements to overlapping regions of $p_T$.


\begin{thebibliography}{9}
\bibitem{ppg054} S.~S.~Adler, {\it et al.} (PHENIX Collaboration), \Journal{\PRC}{76}{034904}{2007}.
\bibitem{QCDcompton} H.~Fritzsch and P.~Minkowski, \Journal{\PLB}{69}{316}{1977}.
\bibitem{ppg086} A.~Adare, {\it et al.} (PHENIX Collaboration), \Journal{\PRL}{104}{132301}{2010}.
\bibitem{Tingpizff} B.~Adeva, {\it et al.} (L3 Collaboration), \Journal{\PLB}{259}{199--208}{1991}.
\bibitem{BW06} N.~Borghini and U.~A.~Wiedemann, \Journal{\NPA}{774}{549--522}{2006}; see also {arXiv:hep-ph/0506218v1}.
\bibitem{ppg095} A.~Adare, {\it et al.} PHENIX Collaboration, \Journal{\PRD}{82}{072001}{2010}.
\bibitem{ConnorsHP2010} M.~Connors, {\it et al.} (PHENIX Collaboration), \Journal{\NPA}{855}{335--338}{2011}. 
\bibitem{LaiHP2010} Y.~S.~Lai, {\it et al.} (PHENIX Collaboration), \Journal{\NPA}{855}{295--298}{2011}. 
\bibitem{ppg029} S.~S.~Adler, {\it et al.} (PHENIX Collaboration), \Journal{\PRD}{74}{072002}{2006}.
\bibitem{ppg106} A.~Adare, {\it et al.} (PHENIX Collaboration), \Journal{\PRL}{104}{252301}{2010}.
\bibitem{MJT-Utrecht} M.~J.~Tannenbaum, {\it et al.} (PHENIX Collaboration), {arXiv:1109.0760v1 [nucl-ex]}
\bibitem{CMSdijet} S.~ Chatrchyan, {\it et al.} (CMS Collaboration), \Journal{\PRC}{84}{024906}{2011}
\end{thebibliography}
\end{document}